\documentclass{webofc}
\usepackage[varg]{txfonts}   
\begin{document}
\title{Possible signals of two QCD phase transitions at
NICA-FAIR energies}
%
%
          
\author{K. A. Bugaev\inst{1} \fnsep\thanks{\email{Bugaev@th.physik.uni-frankfurt.de}} 
\and
        A. I. Ivanytskyi\inst{1, 2}
        \and
         V. V. Sagun\inst{1,3}, B. E. Grinyuk\inst{1}, D. O. Savchenko\inst{1}, 
          G. M. Zinovjev\inst{1} 
        \and
        E. G. Nikonov\inst{4}
        \and
              L. V.  Bravina\inst{5}
        \and
        E. E.  Zabrodin\inst{5, 6, 7}
        \and 
         D. B.  Blaschke\inst{7, 8, 9}
         \and
         S. Kabana\inst{10, 11}
        \and
        A. V.  Taranenko\inst{7}
}

\institute{Bogolyubov Institute for Theoretical Physics of the National Academy of Sciences of Ukraine, 03680 Kiev,  Ukraine
\and
Department of Fundamental Physics, University of Salamanca, 
		Plaza de la Merced s/n 37008, Spain
		\and
Centro de Astrof\'{\i}sica e Gravita\c c\~ao  - CENTRA,
Departamento de F\'{\i}sica, Instituto Superior T\'ecnico,
Universidade de Lisboa, 1049-001 Lisboa, Portugal
\and
 Laboratory for Information Technologies, Joint Institute for Nuclear Research,   Dubna 141980, Russia
 \and
Department of Physics, University of Oslo, PB 1048 Blindern, N-0316 Oslo, Norway 
\and
Skobeltzyn Institute of Nuclear Physics, Moscow State University, 119899 Moscow, Russia
\and
National Research Nuclear University ``MEPhI'' (Moscow Engineering Physics Institute), 115409 Moscow, Russia
\and
Institute of Theoretical Physics, University of Wroclaw, pl. M. Borna 9, 50-204 Wroclaw, Poland
\and
Bogoliubov Laboratory of Theoretical Physics, JINR Dubna, Joliot-Curie str. 6, 141980 Dubna, Russia
\and
University of Nantes,  Facult\'e des Sciences et des Techniques,
2, rue de la Houssini\'ere, 44322 Nantes, France
\and
 SUBATECH, Ecole des Mines, 4 rue Alfred Kastler, 44307 Nantes, France
}

\abstract{%
The chemical freeze-out irregularities found with the most advanced  hadron resonance gas model and 
possible signals of two QCD phase transitions are discussed. 
We found that the center-of-mass collision energy range of tricritical endpoint of QCD phase diagram is [9; 9.2] GeV  which is consistent both with QCD inspired exactly solvable model and with experimental findings.  
  }
\maketitle
\section{Introduction}
\label{Bugaev_intro}

The experimental programs of heavy ion collisions which are planned at  RHIC (BNL) \cite{KAB_RHIC17},  NICA (JINR) \cite{KAB_NICA} and FAIR (GSI) \cite{KAB_FAIR} are aimed to study the phase diagram of strongly interacting matter. 
The major tasks of these experiments are a determination of the threshold collision energies of  two  phase transitions (PTs) expected to exist in  quantum chromodynamics (QCD) and a location of  their tricritical (or two critical endpoints). However, until recently the situation with the reliable 
signals of the chiral symmetry restoration (CSR) and the deconfinement PTs was not clear. Thus, it was not clear whether 
at high baryonic densities there are two different phase transitions  or a single one. Furthermore, neither the order of these transitions nor their
order were clear until recently. 

Although some promising  irregularities in 
the behavior  of experimental data,  known in the literature as 
the Kink \cite{KAB_Kink}, the Strangeness Horn \cite{KAB_Horn} and the Step 
\cite{KAB_Step}, were found, their relation to any of two QCD phase transitions remains  unclear.
There are two  problems with the theoretical model with  which  the collision energy dependence of $K^+/\pi^+$ ratio was predicted  \cite{KAB_Horn}: first, at high baryonic densities the deconfinement PT occurs  in this model at temperature above 200 MeV, which 
simply contradicts to the present lattice QCD findings; second,  within the model of Ref. \cite{KAB_Horn} the number of non-strange degrees of freedom is 16 and the number of strange degrees of freedom is 14, which means that the total number of degrees of freedom is 30 which is essentially below the number of degrees of freedom of QCD with 3 quark flavors and 3 colors. Hence, we consider the model of Ref. \cite{KAB_Horn} as a successful parameterization of experimental data, but one cannot draw a reliable physical conclusions from it. 

At the same time during last few years two theoretical groups, namely the Kiev and the Giessen ones, reported the signals of two  QCD PTs, 
and, remarkably, they independently concluded \cite{KAB_Kiev1, KAB_Kiev2, KAB_Kiev3,  KAB_Giess_1, KAB_Giess_2, KAB_Kiev4} that the  CSR PT occurs in  the hadronic phase at the center-of-mass collision energy 
$\sqrt{s_{NN}}\simeq 4-5$ GeV, 
while the deconfinement PT to quark gluon plasma (QGP) occurs at the collision energy  $\sqrt{s_{NN}}\simeq 9-10$ GeV.
 Furthermore, both groups argued that the CSR is the first order PT \cite{KAB_Giess_1, KAB_Giess_2, KAB_Kiev4},
while the deconfinement PT has  a second order  \cite{KAB_Giess_1, KAB_Giess_2,KAB_Kiev4,SKPM1}  or a weak first order  \cite{KAB_Kiev4}.
Moreover, combining the results of Ref. \cite{KAB_Ko} on the enhancement of light nuclei fluctuations at $\sqrt{s_{NN}}\simeq 8.8$ GeV
with the hypothesis  of existence of two QCD PTs, we came to a conclusion that the collision energy range $\sqrt{s_{NN}}\simeq 8.8-9.2$ GeV
corresponds not to a vicinity of  CEP as suggested in \cite{KAB_Ko}, but to a 3CEP \cite{KAB_Kiev4,KAB_Kiev5}, thus supporting 
\cite{Andronic:2009gj}.

Due to importance of these findings for the planned experimental programs, 
here we would like to discuss them in some details and to reinforce them with the new arguments. 
The work is organized as follows: in Sect. \ref{Bugaev_sect2} the main results of the multicomponent hadron resonance gas model (HRGM) 
are presented; while Sect.  \ref{Bugaev_sect3} is devoted to a discussion of two QCD PTs; our conclusions are summarized in Sect.  \ref{Bugaev_sect4}.

 \section{HRGM with  Multicomponent Hard-Core Repulsion}
 \label{Bugaev_sect2}
 
The possible PTs signals mentioned above were obtained by  Kiev group  with the help of the  multicomponent HRGM  (MHRGM)
\cite{KAB_Kiev1,KAB_Kiev2,KAB_Kiev3,KAB_Kiev4,KAB_Kiev5, KAB_Oliinychenko:12,KABBugaev:12a,KABSFO,KABVeta14, KAB_IST1, KAB_IST2} which has 4 or 5 hard-core radii  of hadrons.
In contrast to 
the HRGM with one or two hard core radii of hadrons \cite{KABAndronic:05}, 
the MHRGM2  of Refs. \cite{KAB_IST1,KAB_IST2}
 has the following hard-core radii  of pions
$R_{\pi}$=0.15 fm, kaons $R_{K}$=0.395 fm,  $\Lambda$-hyperons $R_{\Lambda}$=0.085 fm, other
baryons $R_{b}$=0.365 fm and  other mesons $R_{m}$=0.42 fm.  
Having only 2 or 3 additional   global fitting parameters compared 
to the usual  HRGM  \cite{KABAndronic:05},  one can  achieve  extremely good description of the 
hadronic multiplicity ratios measured at AGS, SPS, RHIC  and LHC energies with a  high quality $\chi^2/dof  \simeq 1.08$  
\cite{KAB_IST1,KAB_IST2}, including traditionally  the 
most problematic ones for the usual HRGM  \cite{KABAndronic:05}, i.e.  $K^+/\pi^+$, $\Lambda/\pi^+$ and $\bar \Lambda/\pi^-$ ratios.
The most successful MHRGM (MHRGM1) based on  the Van der Waals approximation is developed in  \cite{KABVeta14} with
the hard-core radii of baryons $R_b$ = 0.355 fm, mesons $R_m = 0.4$ fm, pions $R_\pi = 0.1$ fm, kaons $R_K = 0.38$ fm, and 
$\Lambda$ hyperons $R_\Lambda = 0.11$ fm which provide a similar fit quality as the MHRGM2.

A high quality fit of hadronic multiplicity ratios achieved by  the MHRGM gives us a high confidence 
that the equation of state  of hadronic matter is now fixed with high accuracy
in the wide range of  chemical freeze-out (CFO) temperature $T$ and baryonic chemical potential $\mu_B$, since
the MHRGM2  \cite{KAB_IST1,KAB_IST2}  allows one to  go beyond the Van der Waals approximation traditionally used in HRGM.

\begin{figure}[t]
\centering
\centerline{\includegraphics[width=63mm,clip]{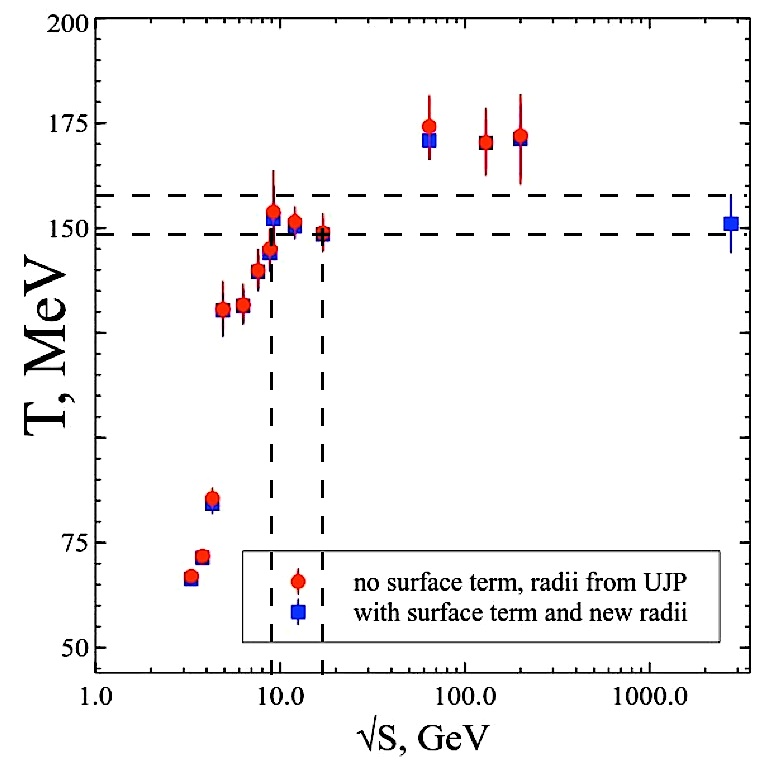}\hspace*{2.2mm}
\includegraphics[width=61.5mm,clip]{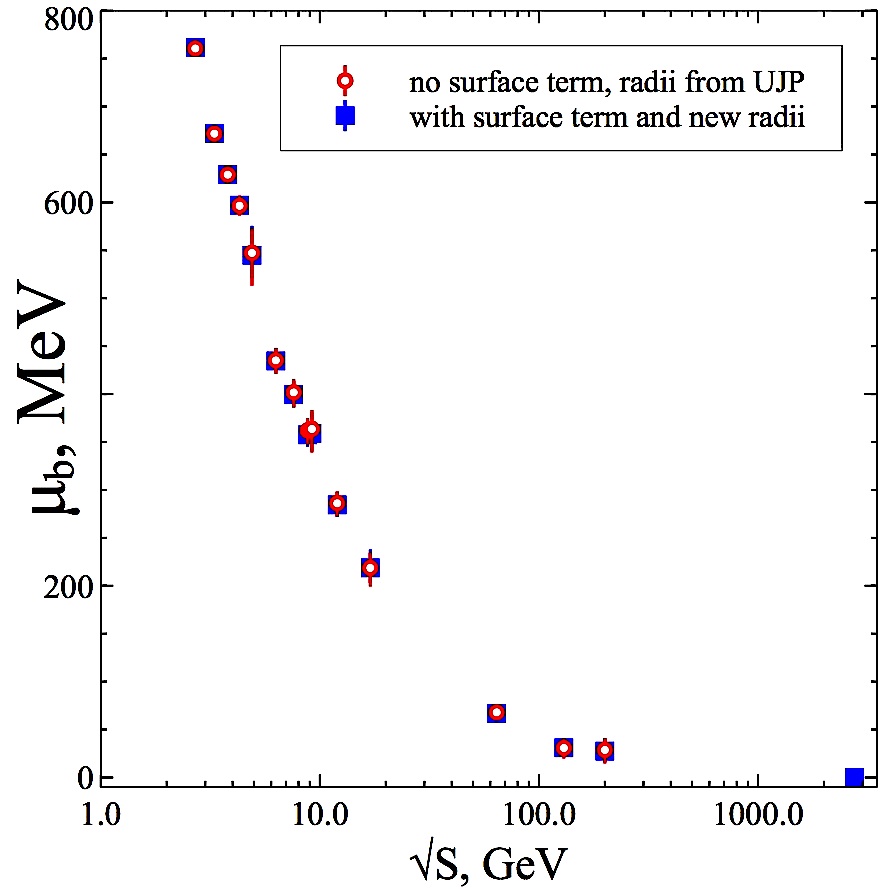}}
\caption{{\bf Left panel:} The CFO  temperature as a function of the collision energy found in \cite{KAB_IST2} within MHRGM1 (circles) and  MHRGM2 (squares). The horizontal dashed lines correspond to the range  of  the (3)CEP temperature values  found in \cite{KAB_Tcep}.
The vertical dashed lines indicate the corresponding values of the collision energy at which the CEP can be reached (see text for details).
{\bf Right panel:} Same as in the left panel, but for the baryonic chemical potential.} 
\label{KAB_fig_T3cep}       
\end{figure}

\begin{figure}[t]
\centering
\centerline{\includegraphics[width=55mm,clip]{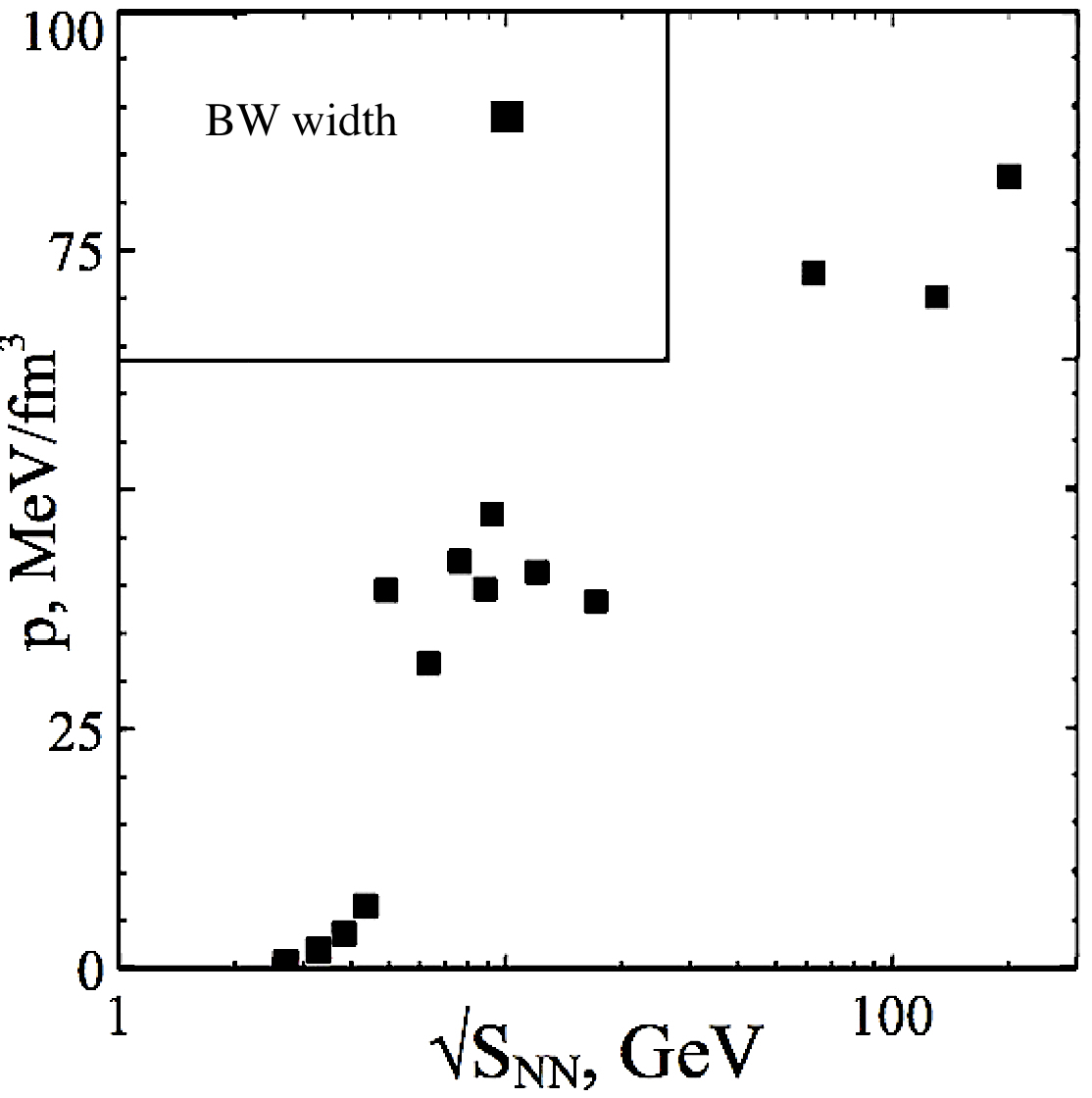}\hspace*{2.2mm}
\includegraphics[width=74.5mm,clip]{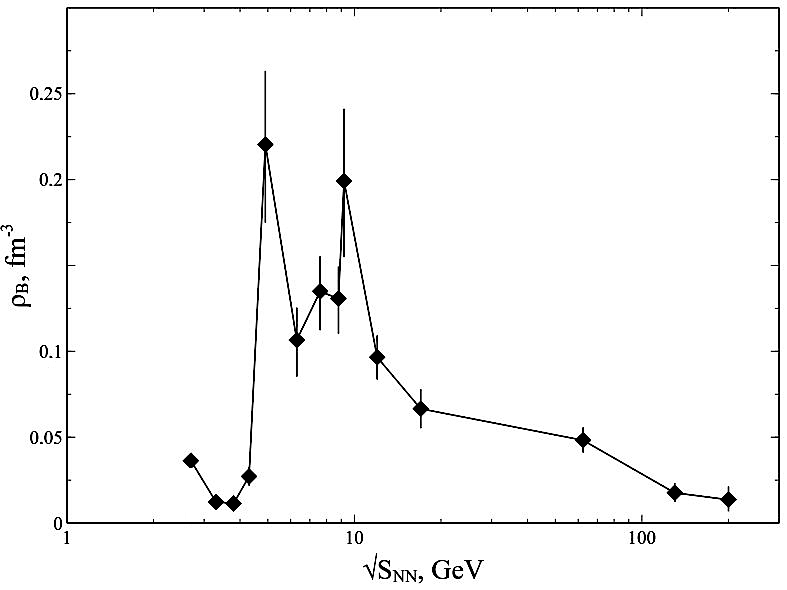}}
\caption{{\bf Left panel:} Pressure at chemical freeze-out  as a function of the collision energy MHRGM2 (squares) \cite{KAB_IST2}.
The  Breit-Wigner parameterization of the resonance width is used in this model. 
{\bf Right panel:} Same as in the left panel, but for the baryonic charge density.} 
\label{KAB_fig_press}       
\end{figure}

More details on the MHRGM and the used fitting procedure can be found in \cite{KABVeta14,KAB_IST1,KAB_IST2,KAB_Kiev5}.
Some results obtained within the MHRGM1 and MHRGM2 are shown in Figs. \ref{KAB_fig_T3cep} and  \ref{KAB_fig_press}. 
From  the left panel of Fig. \ref{KAB_fig_T3cep} one can see that the CFO temperature strongly jumps, when 
the 
collision energy  increase from $\sqrt{s_{NN}}\simeq 4.3$ GeV to $\sqrt{s_{NN}}\simeq 4.9$ GeV. As one can see from the left panel
of Fig.  \ref{KAB_fig_press}, at this energy range  the CFO pressure  increases in 6 times!  The right panel  of Fig.  \ref{KAB_fig_press}
shows that at the collision energy $\sqrt{s_{NN}}\simeq 4.9$ GeV  there is a huge maximum of the baryonic charge density. 
In other words,  increasing the collision energy from $\sqrt{s_{NN}}\simeq 4.3$ GeV to $\sqrt{s_{NN}}\simeq 4.9$ GeV by 16 \%
one can create a dense and small system with very high pressure. The question is what kind of state is it? 

Further analysis of  thermodynamics at CFO performed in works  \cite{KAB_Kiev1, KAB_Kiev2, KAB_Kiev3, KAB_Kiev4}   revealed  a few 
irregularities   observed at CFO  which are
related to two QCD  phase transitions. The most remarkable 
irregularities include two sets of correlated quasi-plateaus  found 
in the collision energy dependence
of the entropy per baryon, total pion number per
baryon, and thermal pion number per baryon
\cite{KAB_Kiev1, KAB_Kiev2, KAB_Kiev3}.  They  are located at the  
collision energy ranges $\sqrt{s_{NN}} \simeq 3.8-4.9$~GeV  and  
$\sqrt{s_{NN}} \simeq  7.6-9.2$~GeV. Also there are  two  peaks of trace anomaly 
$\delta = \frac{(\epsilon - 3p)}{T^4}$ (here $\epsilon$, $p$ and $T$ 
denote, respectively, the energy density of the system, its pressure 
and temperature) observed at the maximal energy of each set of 
quasi-plateaus \cite{KAB_Kiev3,KAB_Kiev4}. 
Also it is remarkable  that  at CFO 
two sharp peaks of  the baryonic charge  density  (see the right panel of Fig. \ref{KAB_fig_press})  are located exactly at the collision energies of the
trace anomaly peaks \cite{KAB_Kiev4}, i.e. at $\sqrt{s_{NN}}=4.9$~GeV 
and $\sqrt{s_{NN}} = 9.2$~GeV. 
This is of outstanding importance, since 
the inflection point/maximum of the trace anomaly is traditionally used in lattice QCD to determine 
the pseudocritical temperature of the cross-over transition \cite{LQCD}. 

Note that the set of low energy   quasi-plateaus
 in the collision energy dependence
of the entropy per baryon, total pion number per
baryon, and thermal pion number per baryon
 was predicted a long time 
ago \cite{KAB:89a,KAB:91} as a manifestation of  the  anomalous 
thermodynamic properties of  mixed phase of first order phase transition.
Since each set of found quasi-plateaus  is  accompanied by the peak of trace anomaly 
and by the peak of baryonic charge density,  in Refs. \cite{KAB_Kiev2,KAB_Kiev3, KAB_Kiev4}
it was inferred  that the high energy quasi-plateaus correspond to another PT. 
However, until very recently the nature of the matter created at the collision
energy range $\sqrt{s_{NN}} \simeq 4.9-9.2$ GeV was unclear. 

 \section{Hadronic Matter with CSR}
 \label{Bugaev_sect3}
 
 Using the MHRGM1 of  Refs. \cite{KAB_Kiev1,KAB_Kiev2} it was possible 
 from fitting the entropy per baryon  
 along the  shock adiabat \cite{KAB:89a,KAB:91} to determine the  equation  of state 
 of the phase created  at 
 the collision energy range $\sqrt{s_{NN}} \simeq 4.9-9.2$ GeV, i.e. between two maxima of baryonic density at CFO (see the right panel of Fig. \ref{KAB_fig_press}).  This   equation of state  is   similar  to the MIT-Bag model
\begin{eqnarray}\label{EqIV}
p_{Chiral} =A_0T^4+A_2T^2\mu^2+A_4\mu^4-B  \,, 
\end{eqnarray}
but the coefficients  $A_0 \simeq 2.53 \cdot 10^{-5}~{\rm MeV}^{-3}{\rm fm}^{-3}$, $A_2 
\simeq 1.51 \cdot 10^{-6} ~{\rm MeV}^{-3}{\rm fm}^{-3}$, $A_4 \simeq 1.001 \cdot 10^{-9}~{\rm 
MeV}^{-3}{\rm fm}^{-3}$, and $B \simeq 9488~{\rm MeV}~{\rm fm}^{-3}$ are rather different 
from what is predicted by the perturbative QCD for massless gluons and  (anti)quarks.  In Ref. \cite{KAB_Kiev4} the equation of state  (\ref{EqIV}) was suggested  to   find  out  the number of  bosonic and fermionic degrees of freedom of this phase.  
Recalling that  first three terms on the right hand side  of Eq.   (\ref{EqIV}) correspond to the gas of  massless particles and noting that the coefficient  $A_4 $ is very small and its value is comparable to its own error, we could determine the numbers of total $N_{dof}^{tot}$,  bosonic $N_{b}^{eff}$  and fermionic 
$N_{f}^{eff}$ degrees of freedom as 
\begin{eqnarray}\label{EqV}
\hspace*{-7mm}N_{dof}^{tot} =  \frac{90}{\pi^2} \, A_0 \hbar^3  \simeq 1770   \,, \quad  N_{f}^{eff} =  12 \, A_2 \hbar^3  \simeq  141 \,, \quad  N_{b}^{eff} =
N_{dof}^{tot} -  \frac{7}{4} N_{f}^{eff} \simeq  1523 \,. 
\end{eqnarray}
Since the numbers  $N_{b}^{eff}$ and $N_{f}^{eff}$ are essentially larger than the corresponding number of degrees of freedom in perturbative QCD, but at the same time  $N_{dof}^{tot}$ is  of  same order as the  total number of spin-isospin degeneracies of all
known non-strange hadrons, 
in Ref.  \cite{KAB_Kiev4} it was  concluded that the equation of state  (\ref{EqIV}) corresponds to the gas of massless hadrons with strong attraction given by the vacuum pressure $B$.  Together with the results of Refs.  \cite{KAB_Giess_1,KAB_Giess_2} these are the strongest
arguments in favor of CSR in hadronic phase.  

It is interesting that the model of Ref.  \cite{KAB_Horn} also employs the equation of state  (1) for the non-strange massless 
degrees of freedom of the ``White Matter", while  the number of degrees of freedom of strange particles which have the mass of  kaons is 14. 
Therefore, by the construction the ``White Matter" model  is a mixture of massless non-strange particles and meson-like strange ones 
with  $N_{dof}^{tot}=30$ \cite{KAB_Horn}. 
Note that the number $N_{dof}^{tot}=30$ does not fit into the QCD with 3 colors and 3 quark flavors. 
Moreover,  the deconfinement PT temperature in the model of Ref. \cite{KAB_Horn} is above 200 MeV which is essentially larger 
than contemporary value of the QCD pseudocritical temperature $T_{pc} =156 \pm 6$ MeV found at vanishing baryonic density \cite{LQCD}. 
Therefore, it is quite  possible that, if the authors of  Ref.  \cite{KAB_Horn} used more realistic equation of state for 
the hadronic phase, it would lower  their PT temperature, but would increase the  number of degrees of freedom. 
Thus, the irregularities found  earlier in   \cite{KAB_Kink,KAB_Horn,KAB_Step}, probably, provide the first evidence 
for an existence of the hadronic matter with CSR, but, due to 
oversimplified  hadronic matter description used in \cite{KAB_Horn}, the estimate obtained in \cite{KAB_Horn} 
for  the collision energy threshold is somewhat higher. 

Now we turn  to discuss the collision energy threshold of the 3CEP. As it was mentioned above 
 the recent analysis of fluctuations of light nuclei  led to a  conclusion that the QCD phase diagram CEP is located in the vicinity
 of  $\sqrt{s_{NN}} \simeq 8.8$ GeV \cite{KAB_Ko,KAB_Ko2}. However, the equation of state used in  Refs. \cite{KAB_Ko,KAB_Ko2} 
 contains the metastable branch (spinodal instability), which is hardly consistent with the hypothesis of local thermal equilibrium
 used in QCD phenomenology.  Hence, we  agree only with the arguments of Refs.  \cite{KAB_Ko,KAB_Ko2}  about the enhancement
 of fluctuations in the vicinity of CEP, but  the signals of two QCD PTs discussed above evidence either of the 3CEP existence \cite{KAB_Kiev4} at which the CSR and deconfinement PTs are matched, or of  existence of two CEPs  for these PTs. 
 The latter situation is, however,  unrealistic, since at low baryonic densities the CSR and deconfinement  are matched 
 in a cross-over. Hence, we believe that the collision energy $\sqrt{s_{NN}} \simeq 8.8$ GeV is a vicinity of 3CEP.
 
This value of  the collision energy is supported by the estimate of the CEP temperature $T_{cep} \in [148.4; 157.4]$ MeV obtained in \cite{KAB_Tcep} on the basis of exactly solvable model  of QGP bags with surface tension. This range of $T_{cep}$ values is shown in the left panel of Fig. \ref{KAB_fig_T3cep} by the horizontal dashed lines. Apparently, the CFO cannot occur above $T_{cep}$ and, hence,  
the CFO temperature of the 3CEP vicinity  obeys the inequality $T_{CFO} \le T_{cep}$. But from the left panel of Fig. \ref{KAB_fig_T3cep} 
it is clear that the collision energy range of the 3CEP vicinity  is $\sqrt{s_{NN}} \in [9; 17.3]$ GeV. Combining these values with
the collision energy range for the deconfinement PT  $\sqrt{s_{NN}} \in [8.8; 9.2]$ GeV obtained recently  in \cite{KAB_Kiev4,KAB_Kiev5},
one finds a narrow range of collision energy for the 3CEP vicinity $\sqrt{s_{NN}} \in [9; 9.2]$ GeV, which simultaneously agrees with
the QCD inspired exactly solvable model and with the experimental signals of deconfinement. 

\section{Conclusions}
 \label{Bugaev_sect4}
 
Here we discuss the MHRGM, the  irregularities found at chemical freeze-out with its help and possible signals of two QCD phase transitions. 
From  the 3CEP  temperature $T_{cep} \in [148.4; 157.4]$ MeV obtained earlier  on the basis of exactly solvable model  of QGP bags with surface tension we found the collision energy range of 3CEP  as $\sqrt{s_{NN}} \in [9; 17.3]$ GeV. However, combining it with 
the collision energy range for the deconfinement PT  $\sqrt{s_{NN}} \in [8.8; 9.2]$ GeV found recently  in \cite{KAB_Kiev4,KAB_Kiev5},
we determine   a narrow range of collision energy for the 3CEP vicinity $\sqrt{s_{NN}} \in [9; 9.2]$ GeV. Evidently, this range is consistent 
both with the QCD inspired model and with experimental findings. 

Comparing the equation of state of  the ``White Matter" model  \cite{KAB_Horn} with the one used to describe the CSR in hadronic matter,
we came to a conclusion that results of Ref. \cite{KAB_Horn} and the associated  irregularities found in  \cite{KAB_Kink,KAB_Step} 
can be considered as a first evidence for the  CSR in hadronic matter. However, the oversimplified parameterization of hadronic phase 
used in  Ref. \cite{KAB_Horn}  led to a larger value of the threshold energy. 

\vspace*{2.2mm}
{\bf Acknowledgments. }{\small  The work of K.A.B., 
A.I.I., V.V.S., B.E.G., D.O.S. and G.M.Z. was supported by 
the grants  launched by the Section of Nuclear Physics of National Academy of Sciences  of Ukraine. 
V.V.S. thanks the 
Funda\c c\~ao para a Ci\^encia e Tecnologia (FCT), Portugal, for the
financial support through the Grant No.
UID/FIS/00099/2013 to make research at the CENTRA, Instituto Superior T\'ecnico, Universidade de 
Lisboa. 
The work of L.V.B. and E.E.Z. was supported by the Norwegian 
Research Council (NFR) under grant No. 255253/F50 - CERN Heavy Ion 
Theory. 
L.V.B. and K.A.B. thank the Norwegian Agency for International Cooperation and Quality Enhancement in Higher Education for financial support, grant 150400-212051-120000 «CPEA-LT-2016/10094 From Strong Interacting Matter to Dark Matter». 
D.B.B. acknowledges a support from the Polish National Science Centre (NCN) under grant no. UMO-2014/15/B/ST2/03752.
The work of A.V.T. was partially supported by the Ministry of Science and Education of the Russian Federation, grant No. 3.3380.2017/4.6.
D.B.B. and A.V.T. acknowledge partial support from the National Research Nuclear University ``MEPhI'' in the framework of the Russian Academic Excellence Project (contract no. 02.a03.21.0005, 27.08.2013).  
The work of A.I.I. was done within the project SA083P17 of Universidad de Salamanca launched by the Regional Government of Castilla y Leon and the European Regional Development Fund. 
The authors are grateful to the COST Action CA15213 ``THOR" for supporting their networking activities. 
}


\begin{thebibliography}{}
%


\bibitem{KAB_RHIC17}
%
L. Adamczyk {et al.,} [STAR Collaboration],
Phys. Rev. C {\bf 93},  021903 (2016).

\bibitem{KAB_NICA}
%
P.  Senger,
{Eur. Phys. J. A} {\bf 52},  217 (2016) and references therein.

\bibitem{KAB_FAIR}
%
P.  Senger,
{Nucl. Phys. A} {\bf 862-863},  139  (2011) and references therein.

\bibitem{KAB_Kink}
M. Gazdzicki, {Z. Phys. C} {\bf 66}, {659} (1995).


\bibitem{KAB_Horn}
%
M. Gazdzicki  and  M.I. Gorenstein, 
{Acta Phys. Polon. B}
{\bf 30}, {2705} (1999).

\bibitem{KAB_Step}
%
M. I. Gorenstein, M. Gazdzicki and K. A. Bugaev,
Phys. Lett. B {\bf 567}, 175  (2003).

\bibitem{KAB_Kiev1}
  K.~A.~Bugaev {et al.,}
 Phys. Part. Nucl. Lett. {\bf 12}, 238  (2015).

\bibitem{KAB_Kiev2}
  K.~A.~Bugaev  {et al.,}
  Eur. Phys. J. A {\bf 52},  175  (2016).

\bibitem{KAB_Kiev3} 
 K. A. Bugaev  {et al.,}
Eur. Phys. J. A {\bf 52}, 227  (2016).


\bibitem{KAB_Giess_1}
%
W. Cassing, A. Palmese, P.  Moreau and E. L. Bratkovskaya,
 Phys. Rev. C   {\bf 93}, 014902 (2016).

\bibitem{KAB_Giess_2}
%
A. Palmese {et al.,} Phys. Rev. C {\bf 94},  044912 (2016).


\bibitem{KAB_Kiev4}
%
  K.~A.~Bugaev
 { et al.,}
 Phys. Part. Nucl. Lett. {\bf  15}, 210 (2018).
  
\bibitem{SKPM1}
S. Kabana and P. Minkowski, Int. J. Mod. Phys. A {\bf 26}, 3035 (2011).

\bibitem{KAB_Ko}
K. J. Sun, L. W. Chen, C. M. Ko, and Z. Xu, 
Phys. Lett. B {\bf 774}, 103 (2017).

\bibitem{KAB_Kiev5}
%
  K.~A.~Bugaev
 { et al.,}
KnE Energ.\ Phys.\  {\bf 3},  313 (2018).
  
\bibitem{Andronic:2009gj} 
  A.~Andronic, D. Blaschke et al.,
  Nucl.\ Phys.\ A {\bf 837}, 65 (2010).
  
\bibitem{KAB_Oliinychenko:12}
%
D. R. Oliinychenko, K. A. Bugaev and A. S. Sorin, 
Ukr. J. Phys.
{\bf 58}, 211  (2013).
 
\bibitem{KABBugaev:12a}
%
K. A. Bugaev, D. R. Oliinychenko,   A. S. Sorin and G. M. Zinovjev, 
Eur. Phys. J. A {\bf 49}, 30 (2013).  

\bibitem{KABSFO}
K. A. Bugaev et al., 
{Europhys. Lett.}  {\bf 104}  22002 (2013)  and references therein.   
   
\bibitem{KABVeta14}
V. V. Sagun,   {Ukr. J. Phys.} {\bf 59} 755 (2014).

\bibitem{KAB_IST1}
%
K. A. Bugaev  et al., 
Nucl. Phys. A {\bf 970},  133 (2018).

\bibitem{KAB_IST2}
%
V. V. Sagun et al., 
 Eur. Phys. J. A {\bf 54},   100  (2018).

\bibitem{KABAndronic:05}
%
A. Andronic, P. Braun-Munzinger   and  J. Stachel
{Nucl. Phys. A}
{\bf 772}, 
{167}  (2006).

\bibitem{KAB_Tcep}
%
K. A. Bugaev et al., 
{Phys. Atom. Nucl.} {\bf 75},  707 (2012).

\bibitem{KAB:89a}
%
 K. A. Bugaev, M. I. Gorenstein,  B.  K\"ampfer and V. I. Zhdanov, 
{Phys. Rev. D}
{\bf 40},
{2903} (1989).

\bibitem{KAB:91}
%
 K. A. Bugaev, M. I. Gorenstein   and D. H. Rischke,  
{Phys. Lett. B}
{\bf 255}, {18} (1991).

\bibitem{LQCD}
%
Sz. Borsanyi et al., JHEP {\bf 08}, 053 (2012) and references therein.



\bibitem{KAB_Ko2}
%
K. J. Sun, L. W. Chen, C. M. Ko,  J. Pu and Z. Xu, 
arXiv:1801.09382v1 [nucl-th].

\end{thebibliography}
%
%

\end{document}